\documentclass[10pt]{iopart}

\usepackage{amssymb,amsmath,mathptmx}
\usepackage{graphicx}
\usepackage{siunitx}
\usepackage{floatrow}
\usepackage{relsize}
\usepackage{gensymb}
\usepackage{xcolor}

\usepackage[square,numbers,sort&compress]{natbib}

\usepackage[breaklinks=true,colorlinks=true,bookmarks=false,urlcolor=blue,
citecolor=blue,linkcolor=blue]{hyperref}

\def\unit#1{\mathord{\thinspace\rm #1}}

\begin{document}

\title{Ferroelectricity in twisted double bilayer graphene}

\author{Renjun Du$^{1,3}$, Jingkuan Xiao$^{1,3}$, Di Zhang$^1$, Xiaofan Cai$^1$, Siqi Jiang$^1$, Fuzhuo Lian$^1$, Kenji Watanabe$^2$, Takashi Taniguchi$^2$, Lei Wang$^{1,*}$ and Geliang Yu$^{1,*}$ }

\address{$^1$ National Laboratory of Solid State Microstructures, School of Physics, Nanjing University, Nanjing 210093, China}
\address{$^2$ National Institute for Material Science, 1-1 Namiki, Tsukuba 305-0044, Japan}
\address{$^3$ These authors contributed equally to this work.}
\address{$^*$ Authors to whom any correspondence should be addressed.}

\ead{leiwang@nju.edu.cn and yugeliang@nju.edu.cn}

\vspace{10pt}
\begin{indented}
\item[]\today
\end{indented}

\vspace{10pt}
\begin{indented}
\item[]{\it Keywords}: Twisted double bilayer graphene, ferroelectricity, moir\'{e} superlattice
\end{indented}

\begin{abstract}

\noindent \normalsize  Two-dimensional ferroelectrics can maintain vertical polarization up to room temperature, and are, therefore, promising for next-generation nonvolatile memories.
Although natural two-dimensional ferroelectrics are few, moir\'{e} superlattices provide us with a generalized method to construct ferroelectrics from non-ferroelectric parent materials. 
We report a realization of ferroelectric hysteresis in a AB-BA stacked twisted double bilayer graphene (TDBG) system.
The ferroelectric polarization is prominent at zero external displacement field and reduces upon increasing displacement fields. 
TDBG in the AB-BA configuration possesses a superlattice of non-centrosymmetric domains, exhibiting alternatively switchable polarities even without the assistance of any boron nitride layers; however, in the AB-AB stacking case, the development of polarized domains necessitates the presence of a second superlattice induced by the adjacent boron nitride layer.
Therefore, twisted multilayer graphene systems offer us a fascinating field to explore two-dimensional ferroelectricity.
\end{abstract}

%


\submitto{\TDM}
%
%
\ioptwocol
%
         
\section{Introduction}
Ferroelectric materials possess a switchable electrical dipole moment that can be reversed by an external electrical field, enabling a wide range of applications in various areas, such as transducers,  pyroelectric sensors, electrocaloric heat pumps, non-volatile memories and photovoltaic devices~\cite{Martin2016Thin}.
Its potential use in next-generation transistors with dense storage and low-power consumption instigates rapid developments in two-dimensional (2D) ferroelectrics.
Despite thin-film ferroelectrics, van der Waals layered materials provide promising candidates for realizing 2D ferroelectrics with atomic thickness and intriguing properties~\cite{Wang2023Towards, Zhang2023Ferroelectric}. 

Natural van der Waals ferroelectrics with polar space groups are few~\cite{Liu2016room, Fei2018ferro,Yuan2019MoTe, Cui2018inter,Zhou2017out, Xiao2020Berry}, however constructing non-centrosymmetric structures with moir\'{e} superlattices open up new possibilities for designing unconventional ferroelectrics, expanding the family of 2D ferroelectric materials~\cite{Weston2022inter, Magorrian2021multi, Enaldiev_2021piezo, Li2017binary, Zhao2021lateral, Zhang2020_moire}.
An array of domains with periodically alternating polarization can be created in moir\'{e} superlattices~\cite{Enaldiev2022scalable}.
Applying an external electrical field can manipulate the polarity of domains, allowing the occurrence of ferroelectricity.
This effect has been demonstrated in a variety of van der Waals materials, such as parallel-stacked hexagonal boron nitride (hBN)~\cite{Yasuda2021stack,Woods2021charge,fraunie2023gatefree, Stern2021interfacial} , twisted transition metal dichalcogenides ( MoSe$_\text{2}$, WSe$_\text{2}$, MoS$_\text{2}$, and WS$_\text{2}$) homo- and hetero-structures~\cite{Wang2022Jan, Weston2022inter, an2023unconventional, Deb2022Cumulative, Rogee2022Ferroelectricity}, and graphene-based supperlattices~\cite{Zheng2020unconvent, Niu2022Giant, Zheng2023Electronic}.    
Interlayer lattice sliding~\cite{Li2017binary, Yang2023Across, Weston2022inter, Stern2021interfacial} is found to well account for ferroelectric phenomena in twisted hBN and transition metal dichalcogenides; nevertheless, it is insufficient to explain the complex behaviour of ferroelectricity in bilayer-graphene-hBN superlattices, namely, layer-specific anomalous screening.
Although the other scenarios considering correlated interactions~\cite{Zhu2022Electric, Zheng2020unconvent, Zheng2023Electronic}  have been proposed, the real nature is still under discussion.
On the other hand, in addition to realizing ferroelectricity in multi-element van der Waals materials, mixed-stacking graphite or twisted multilayer graphene offer us a novel system to achieve single-element ferroelectricity~\cite{Garcia-Ruiz2023mixed, yang2023atypical, atri2023spontaneous, winterer2023ferroelectric}, which requires extensive experimental investigations.

Here, we present an experimental observation of ferroelectricity in a twisted double bilayer graphene system. 
We have constructed AB-BA stacked TDBG, which is anticipated to exhibit ferroelectricity as a result of the existence of symmetry-broken domains with opposite polarity.
In addition, we have discussed the potential impact of the additional superlattice, induced by the adjacent hBN layer, on ferroelectricity.
However, it is important to highlight that such additional superlattice is not a prerequisite for the generation of ferroelectricity in a twisted multilayer graphene system. 
Our experimental findings may improve the understanding of ferroelectricity in the intricate graphene-based superlattice system.

 \begin{figure*}[tb]
	\includegraphics{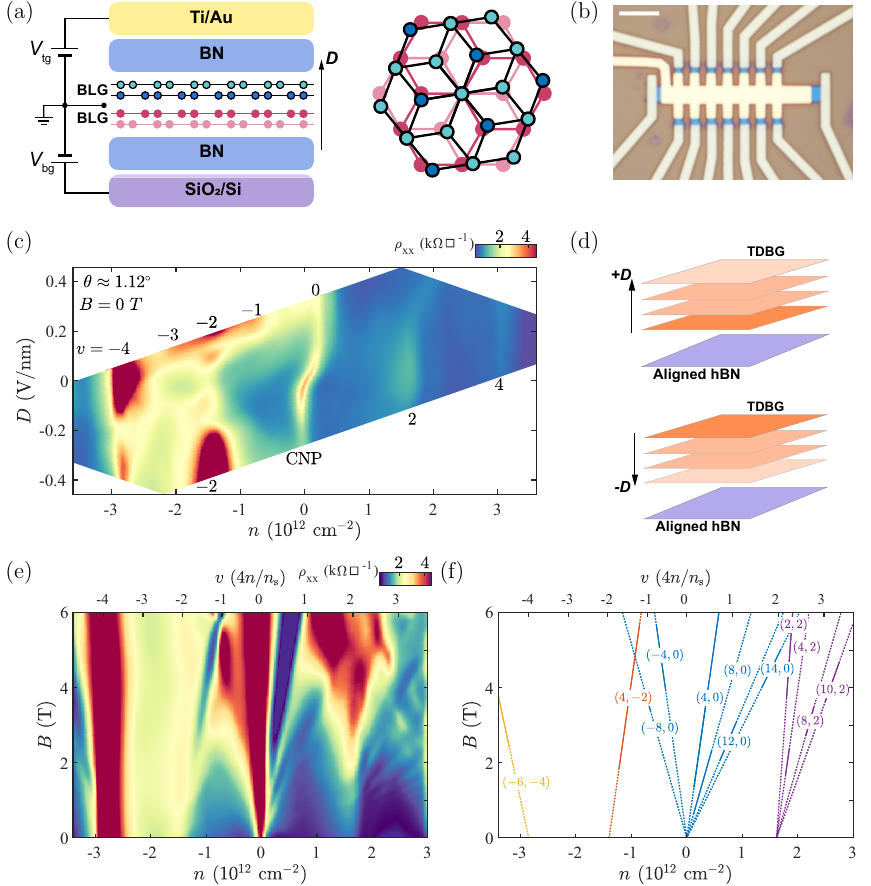}
	\caption{Twisted double bilayer graphene (sample S1). (a) Schematics of our dual-gated device made from twisted AB-BA stacked double bilayer graphene. (b) An optical image of the investigated device. The scale bar is 5\,\si{\mu m}. (c) Longitudinal resistivity measured by tuning charge carrier densities ($n$) and external displacement fields ($D$) at 1.6\,\si{K} and zero magnetic field. Resistive maxima appear at the filling factors $\nu=4n/n_\mathrm{s} = -4, -3, -2, -1, 2, 4$ ($n_\mathrm{s}$ is the electron density of 4 electrons per moir\'{e} unit cell) and the charge neutrality points. (d) Schematics of the layer-resolved hole density for positive and negative $D$.  (e) Landau fan diagram as a function of charge carrier densities and perpendicular magnetic fields at $D = 0$. (f) Schematics of the corresponding gapped states in (e). 
	Different sets of Landau levels are grouped with distinct colors according to their intersections at $B=0$.
		 \label{Fig_1}} 
\end{figure*}
\section{Results and discussions}
We have fabricated a dual-gated device based on AB-BA stacked TDBG, as shown in figure~\ref{Fig_1}(a). 
A \textquoteleft flip-and-stack\textquoteright\ approach~\cite{yang2020in} has been utilized to construct the AB-BA stacked double-bilayer graphene (see details in the supplemental material).
We first picked up half of bilayer graphene (BLG) with a hBN flake affixed to a PDMS/PC (Polydimethylsiloxane/Poly(Bisphenol A carbonate)) stamp~\cite{Kim2016van, cao2016super}, and then prepared the remaining part using the same procedure but with a second PDMS/PC stamp. 
The two stamps were brought into contact face-to-face by flipping one of them over.
The twisted angle was controlled by carefully aligning the crystal axes of BLG under a high precision transfer set-up. 
The entire heterostructure (hBN-BLG-BLG-hBN) was picked up by the upper stamp and placed on a Si/SiO$_\text{2}$ substrate.
The electrical field can be applied through a metallic top gate (Ti/Au) and a Si back gate. 
The sample was designed as a Hall bar and connected by depositing metallic edge contacts (Ti/Al/Au), as shown in figure~\ref{Fig_1}(b).
The device was measured at 1.6 K using a standard low-frequency lock-in technique.

Twisted double-bilayer graphene  presents signatures of correlated interaction  regardless of whether the stacking order is AB-AB or AB-BA~\cite{Cao2020tunable, He2021sym, Burg2019corr, Liu2020tunable, Shen2020corre, he2021chirality, Wang2022emergence, Kuiri2022spontaneous}.  
Figure~\ref{Fig_1}(c) shows the longitudinal resistivity $\rho_\mathrm{xx}$
of an AB-BA stacked TDBG device with a twist angle of $1.12^{\circ}$ as a function of charge carrier densities $n$ and displacement fields $D$.
The charge carrier density is defined as  $n = ( C_\mathrm{bg}V_\mathrm{bg} + C_\mathrm{tg}V_\mathrm{tg}) / e$, where $C_\mathrm{bg}$ ($C_\mathrm{tg}$) is the capacitance between the back (top) gate  and TDBG, and $e$ is the elementary charge; and the external displacement field is obtained by $D = ( C_\mathrm{bg}V_\mathrm{bg} - C_\mathrm{tg}V_\mathrm{tg}) / 2\epsilon_\mathrm{0}$ ($\epsilon_{0}$ is the vacuum permittivity), penetrating upward through the device.
Unlike the previous works~\cite{Cao2020tunable, He2021sym, Burg2019corr, Liu2020tunable, Shen2020corre, he2021chirality, Wang2022emergence, Kuiri2022spontaneous},  there is no sign of a gap opening at the charge neutrality point (CNP), and the resistivity is decreased for higher $D$,  implying the touching between the first conduction and valence bands.
Although complete isolation of the flat bands is not realized, correlated states still appear on both electron and hole sides, resulting in a few resistance maxima at multiples of 1/4 filling of the flat bands. 
In contrast to earlier research~\cite{Cao2020tunable, He2021sym, Burg2019corr, Liu2020tunable, Shen2020corre, he2021chirality, Wang2022emergence, Kuiri2022spontaneous}, features of insulating states only exist on the hole side rather than the electron side, indicating that the bandwidth of the first valence band is narrower than that of the first conduction band.
Usually, correlated states are symmetric for the variation of $D$ in TDBG, but we found discernible asymmetric behaviours in resistivity, especially for $\nu=-1, -2, -3$.   
The occurrence of insulating states at $\nu=-1, -3$ and $D>0.1\unit{V/nm}$ originates from an additional superlattice between BLG and encapsulating BN which causes the nonequivalent dependence of the band structure on the direction of $D$~\cite{he2021chirality}.
For positive $D$, when the field penetrates from the bottom BLG to the top BLG, the hole density in the outer sheet of bottom BLG is higher; while it is suppressed for negative $D$ (see figure~\ref{Fig_1}(d)).
Because the asymmetric features of correlated states at $\nu=-1, -3$ on the positive $D$ side favour high-density states of holes, we can conclude that the bottom BN flake aligns with the bottom BLG layer.
Nevertheless, the rotation angle between the bottom BN and BLG layers is relatively large ($\sim5.6\unit{\degree}$) such that we didn't observe any associated resistivity peaks in transport measurements (see the alignment information in the supplemental material). 

\begin{figure*}[htb]
	\includegraphics{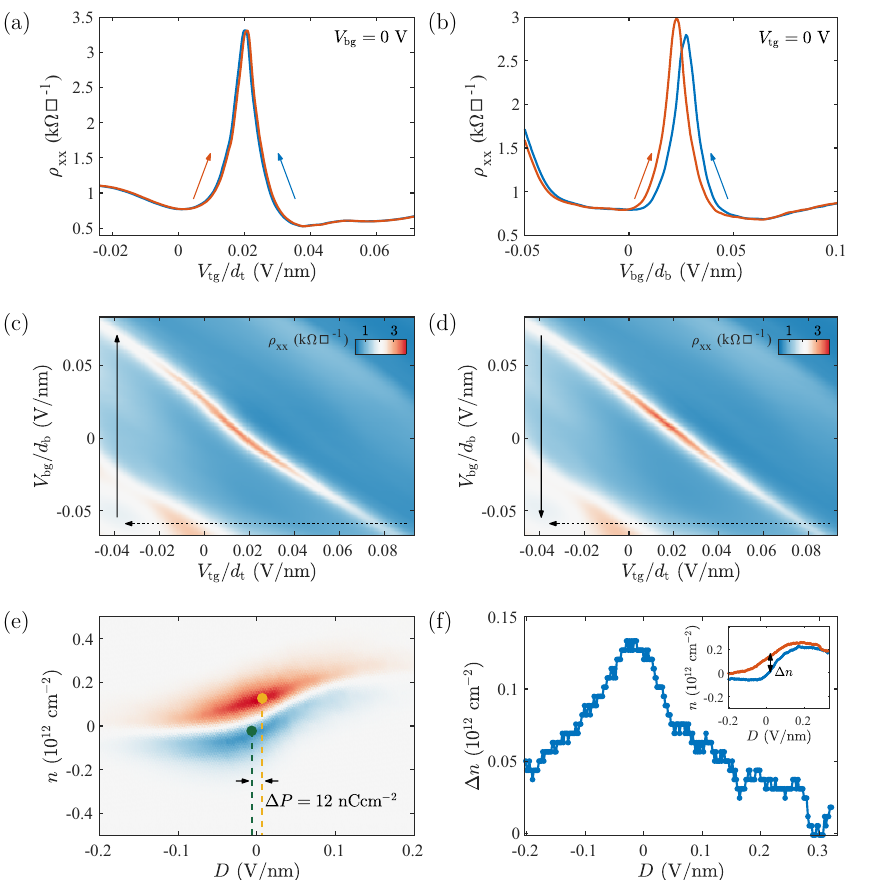}
	\caption{Ferroelectricity in AB-BA stacked twisted double bilayer graphene (sample S1). (a) Longitudinal resistivity shows no hysteresis for the forward (red curve) and backward (blue curve) scans of the top-gate voltage at $V_\mathrm{bg} = 0$. (b) Ferroelectric hysteresis of the longitudinal resistivity occurs during sweeping the back-gate voltage forward (red curve) and backward (blue curve) at $V_\mathrm{tg} = 0$. Here, the $x$-axis $V_\mathrm{tg}$ ($V_\mathrm{bg}$) is normalized by the top (bottom) BN thickness $d_\mathrm{t}$ ($d_\mathrm{b}$).  In (c) and (d), the longitudinal resistivity is tuned by $V_\mathrm{tg}$ and $V_\mathrm{bg}$ together. We use $V_\mathrm{tg}$ as the fast scan axis and $V_\mathrm{bg}$ as the slow scan axis. The dashed and solid arrows denote the scan directions of $V_\mathrm{tg}$ and $V_\mathrm{bg}$, respectively. The forward (c) and backward (d) scans of the back-gate voltage are shown separately. (e) The difference of longitudinal resistivity between the forward (c) and backward (d) resistivity maps, $\Delta \rho_\mathrm{xx} = \rho_\mathrm{xx}^\mathrm{forward} - \rho_\mathrm{xx}^\mathrm{backward}$, as a function of charge carrier densities and external displacement fields. The green and yellow dots indicate two  resistivity maxima of the charge neutrality points for the forward and backward sweeps, respectively. The remnant polarization is defined as $\Delta P = P^\mathrm{forward} - P^\mathrm{backward} = 12\,\si{nC\cdot cm^{-2}}$. (f) The difference in charge carrier densities ($\Delta n$) as a function of external displacement fields. Inset shows the positions of charge neutrality points in (c) and (d) with the red- and blue-dotted curves, respectively. \label{Fig_2}}
\end{figure*}

To explore the topological properties of the flat bands in AB-BA stacked TDBG, we investigated the evolution of resistivity in magnetic fields.
Figure~\ref{Fig_1}(d) shows a number of linearly dispersed Landau levels that are constituents of a Wannier diagram which is described by the Diophantine equation $\nu=t\phi/\phi_{0}+s$, where $\nu$ is the filling factor of electron density per Bloch band, $\phi$ is the magnetic flux per moir\'{e} unit cell, $\phi_{0}$ is the magnetic flux quantum, $t$ represents the quantized Hall conductivity $\sigma_\mathrm{xy} = -te^2/h$, and $s$ is the Bloch band filling at each gap.
At $D=0$, these Landau levels can be classified into two classes using topologically invariant integers ($t$,$s$).
For $s=0,-4$ and $t\neq0$, the Fermi level locates at the CNP or a single particle gap where the magneto-oscillations arise from integer quantum Hall effect.
Around the CNP, the Landau levels take the sequence of $t=\pm4,\pm8,12,14$, presenting fourfold degeneracy as that in twisted bilayer graphene (TBG).
Such lifting of eight-fold degeneracy in TBG may originate from  $C_{3}$ symmetry broken\cite{zhang2019landau, Saito2021Hofstadter,Yu2022Correlated}.
In the case of TDBG, the number of degenerate states may also depend on stacking order.
The AB-BA stacked TDBG lacks valley degeneracy while the AB-AB stacked TDBG preserves it even though these two configurations have nearly the same band structure at zero magnetic field\cite{crosse2020hofstadter}.
Accordingly, the degeneracy of the Landau levels in AB-BA stacked TDBG is only half of that in the AB-AB case under weak magnetic fields. 
In the absence of triangle warping, the fourfold degeneracy in the AB-BA configuration may occur even without considering other symmetry breaking\cite{crosse2020hofstadter}.
On the other hand, states with integer $s=\pm2$ and $t\neq0$ correspond to correlated Chern insulators.
The twofold degeneracy is found at $s=2$ as the case in TBG\cite{Saito2021Hofstadter,Yu2022Correlated}.

Apart from correlated states, the heterostructrue of AB-BA stacked TDBG is also a fascinating system for the realization of ferroelectricity.
A hysteretic behaviour of resistivity can be detected with the control of gate voltages.
When we sweep the top gate forward (red curve) and backward (blue curve) while keeping $V_\mathrm{bg} = 0$, the resistivity shows no lag between the two opposite sweeps (see figure~\ref{Fig_2}(a)).
However, a discernible hysteresis occurs when we tune the back-gate voltage forward and backward at $V_\mathrm{tg} = 0$, implying ferroelectric polarization (see figure~\ref{Fig_2}(b)). 
We use $V_\mathrm{tg}$ and $V_\mathrm{bg}$ as axes for the fast and slow scans, respectively, and then obtain two resistivity maps shown in figures~\ref{Fig_2}(c) and (d) corresponding to the forward and backward sweeps of $V_\mathrm{bg}$, respectively.
The hysteresis is most apparent  along the charge neutrality line that is bending with distinct curvatures for the forward and backward scans.
The difference of resistivity between the forward- and backward-scan maps, \textit{i. e.}, $\Delta \rho_\mathrm{xx} =  \rho_\mathrm{xx}^\mathrm{forward} - \rho_\mathrm{xx}^\mathrm{backward}$, is plotted as a function of $D$ and $n$ in figure~\ref{Fig_2}(e).
The ferroelectric polarization ($P$) generates an internal electrical field against the external field ($D$), leading to the resistivity maximum, originating from the band touching point, shifting from $D=0$ to $D = -16.25\,\si{mV/nm}$ for the forward sweep (green dot) and to $D = 16.25\,\si{mV/nm}$ for the backward scan (yellow dot)~\cite{Zheng2020unconvent}.
The total shift ($\Delta D$) between the two resistivity maxima illustrates the switchable remnant polarization, which can be defined as $\Delta P = \epsilon_{0}\Delta D = 12\,\si{nC\cdot  cm^{-2}}$.
The ferroelectric polarization also causes a change in carrier density ($\Delta n$), leading to a lag of the CNP in $\rho_\mathrm{xx}$ (see figure~\ref{Fig_2}(f)).
$\Delta n$ reduces with increasing $|D|$ and reaches its maximum at $D = 0$, indicating the largest polarization.

\begin{figure*}[htp]
	\includegraphics{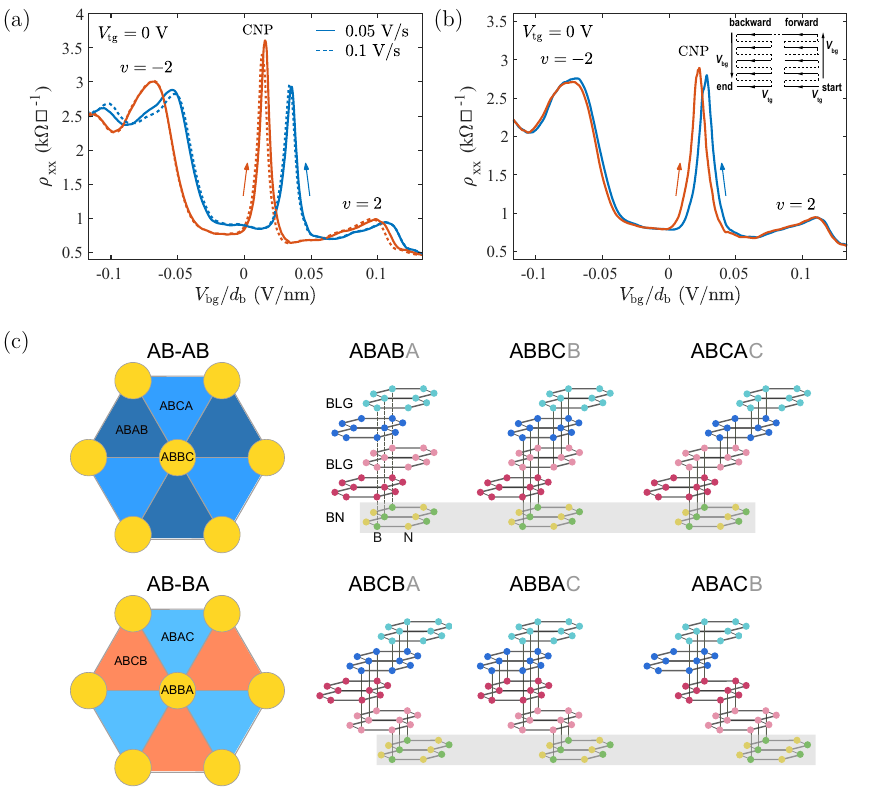}
	\caption{Ferroelectricity affected by the motion of domains (sample S1). (a) Longitudinal resistivity measured by sweeping the back-gate voltage forward (red) and backward (blue) at $V_\mathrm{tg} = 0$. The dashed and solid lines denote scans with the fast- and slow-scan rates, respectively. The hysteresis is shown at both the charge neutrality point and half-filling positions.
		(b) Longitudinal resistivity measured with respect to $V_\mathrm{bg}$ at $V_\mathrm{tg} = 0$ after a sequential manipulation of both top- and back-gate voltages.  The black zigzag lines describe the sequence of the electrical training process. A reduction of hysteresis is observed. 
		(c) Schematics of the stacking order of domains in the AB-AB and AB-BA stacked TDBG superlattices. In the case of AB-AB stacked TDBG, the domains present three types of centrosymmetric stacking order, that is ABAB, ABBC, and ABCA, which can become non-centrosymmetric by adding a superlattice using an adjacent BN lattice (grey-shaded region). 
		The structures for the five-layer domains change to ABAB\textcolor{gray}{A}, ABBC\textcolor{gray}{B} and ABCA\textcolor{gray}{C} with the boron atom sitting below the carbon atom, which is energetically favourable~\cite{Yang2023Across, Moore2021Nanoscale}.
		For the AB-BA stacked TDBG, it shows three non-centrosymmetric domains with ABCB, ABBA, and ABAC stacking, which give rise to a lattice with alternating polarized domains (up-polarized ABCB and down-polarized ABAC). 
		When adding a superlattice with BN, the domains are still non-centrosymmetric, taking the stacking structures of ABCB\textcolor{gray}{A}, ABBA\textcolor{gray}{C}, and ABAC\textcolor{gray}{B}.
		\label{Fig_3}}
\end{figure*}

The motion of domains in moir\'{e} superlattices has been proven to play a significant role in the switch of ferroelectric polarization~\cite{Yasuda2021stack,Woods2021charge,Weston2022inter}.
Here, we present evidence of this effect in AB-BA stacked TDBG.
Figure~\ref{Fig_3}(a) shows resistivity with more significant hysteresis that occurs not only at the CNP but also at the resistive peaks arising from correlated insulating states, implying the coexistence of ferroelectricity and strong correlations.
Different scan rates have been used to demonstrate the stability of this large hysteresis.
However, the hysteretic behaviour is eliminated to a great extent and limited in a small region around the CNP after we perform an electrical training process by tuning $V_\mathrm{bg}$ and $V_\mathrm{tg}$ together, as illustrated in figure~\ref{Fig_3}(b).
The remaining hysteresis is stable during measurements in the following months.
We may ascribe this variation of hysteresis to the motion of domains facilitated by interlayer lattice sliding ~\cite{Woods2021charge,Weston2022inter, Zhang2022Domino}.
When tuning $D$ with both gates, the polarization of domains can be changed, allowing a ferroelectric switch.
If the domain wall is pinned around blisters, wrinkles, or defects,  the switch of ferroelectricity is impeded, resulting in the reduction of hysteresis~\cite{Weston2022inter}.

Interlayer lattice sliding is known to be a driving mechanism for the emergence of ferroelectricity~\cite{Stern2021interfacial, Yang2023Across, Li2017binary, Garcia-Ruiz2023mixed, Enaldiev2022scalable, molino2023Ferroelectric}. 
Creating moir\'{e} superlattices becomes a generalized approach to establish arrays  of ferroelectric domains with switchable polarity.
The structure of domains can be either centrosymmetric or non-centrosymmetric.
For example, the stacking configurations in TDBG, such as ABAB and ABCA, are centrosymmetric and lack polarity; while ABCB and ABAC are non-centrosymmetric and opposite in polarity (see figure~\ref{Fig_3}(c)).
Accordingly, sliding the domains from up-polarized to down-polarized or polarity to non-polarity can generate ferroelectricity.
In the case of AB-BA stacked TDBG, a ferroelectric switch between ABCB and ABAC may exist, leading to the hysteretic behaviour of resistivity even without additional alignment of BN.
Nevertheless, for AB-AB stacked TDBG, ferroelectricity can also be introduced by building an additional moir\'{e} structure using BN (see more data in the supplemental material), similar to the case in the BN-encapsulated bilayer graphene~\cite{Zheng2020unconvent, Niu2022Giant, Zheng2023Electronic}.
By adding the fifth layer of BN to TDBG ( see the grey-shaded regions in figure~\ref{Fig_3}(c)), the domains with all kinds of stacking order become non-centrosymmetric so that AB-AB stacked TDBG can also demonstrate ferroelectricity.
In sample S1 with the AB-BA configuration, although bottom BN is aligned with TDBG by $\sim5.6\unit{\degree}$, its effect on the formation of electrical dipoles might be weakened or even vanished~\cite{Niu2022Giant} due to the small moir\'{e} wavelength of $\sim2.5\unit{nm}$.
Interfacial lattice sliding provides a potential theoretical picture of ferroelectricity in TDBG without layer-specific-anomalous screening, even though distinct explanations~\cite{Zhu2022Electric, Zheng2020unconvent, Zheng2023Electronic}, such as interaction-driven charge transfer, are still under discussion.
\section{Conclusions}
We have demonstrated the observation of ferroelectricity  in AB-BA stacked TDBG assembled with a \textquoteleft flip-and-stack\textquoteright\ transfer method. 
Compared to TDBG with the AB-AB configuration, the AB-BA arranged TDBG exhibits similar band structures but with distinct topological properties, which may develop topologically non-trivial features in the Hofstadter spectrum, such as the lifting of valley degeneracy.
On the other hand, TDBG with the AB-BA stacking also presents ferroelectricity as expected for the non-centrosymmetric domain structures, which is in sharp contrast to TDBG in the AB-AB case.
In addition, the double moir\'{e} superlattices generated by BLG-BLG and BLG-hBN may contribute to the ferroelectric behaviour together by constructing five layer non-centrosymmetric domains.
The role of moir\'{e} potential on ferroelectricity in graphene-based heterostructures is still under discussion. 
Our work provides new evidence which may improve the understanding of this phenomenon.

\ack We thank A. S. Mayorov, G. Ma, Y. Han, P. Wang and J. Huang for their fruitful discussions. Financial support from the National Key R\&D Program of China (Nos. 2018YFA0305804, 2018YFA0306800), the National Natural Science Foundation of China (Nos. 12004173, 11974169), the Natural Science Foundation of Jiangsu Province (Nos. BK20220066), and the Fundamental Research Funds for the Central Universities (Nos. 020414380087, 020414913201) are gratefully acknowledged.


\bibliographystyle{iopart-num}

\bibliography{FE_reference}

\end{document}